\documentclass[pra,showpacs,superscriptaddress,twocolumn]{revtex4-1}

\usepackage[dvipdfmx]{graphicx}
\usepackage{amsmath}
\def\vector#1{\mbox{\boldmath $#1$}} 
\usepackage{color}
\usepackage{cancel}
\usepackage{ulem}
\usepackage{comment} 
\usepackage{bm} 

\begin{document}
\title{Helicity-Changing Brillouin Light Scattering by Magnons in a Ferromagnetic Crystal}

\author{R.~Hisatomi}
\email{ryusuke.hisatomi@qc.rcast.u-tokyo.ac.jp}
\affiliation{Research Center for Advanced Science and Technology (RCAST), The University of Tokyo, Meguro-ku, Tokyo 153-8904, Japan}
\author{A.~Noguchi}
\affiliation{Komaba Institute for Science (KIS), The University of Tokyo, Meguro-ku, Tokyo 153-8904, Japan}
\author{R.~Yamazaki}
\affiliation{Research Center for Advanced Science and Technology (RCAST), The University of Tokyo, Meguro-ku, Tokyo 153-8904, Japan}
\author{Y.~Nakata}
\affiliation{Research Center for Advanced Science and Technology (RCAST), The University of Tokyo, Meguro-ku, Tokyo 153-8904, Japan}
\author{A.~Gloppe}
\affiliation{Research Center for Advanced Science and Technology (RCAST), The University of Tokyo, Meguro-ku, Tokyo 153-8904, Japan}
\author{Y.~Nakamura}
\affiliation{Research Center for Advanced Science and Technology (RCAST), The University of Tokyo, Meguro-ku, Tokyo 153-8904, Japan}
\affiliation{Center for Emergent Matter Science (CEMS), RIKEN, Wako, Saitama 351-0198, Japan}
\author{K.~Usami}
\email{usami@qc.rcast.u-tokyo.ac.jp}
\affiliation{Research Center for Advanced Science and Technology (RCAST), The University of Tokyo, Meguro-ku, Tokyo 153-8904, Japan}

\date{\today}

\begin{abstract}
Brillouin light scattering in ferromagnetic materials usually involves one magnon and two photons and their total angular momentum is conserved. Here, we experimentally demonstrate the presence of a helicity-changing \textit{two-magnon} Brillouin light scattering in a ferromagnetic crystal, which can be viewed as a four-wave mixing process involving two magnons and two photons. Moreover, we observe an unconventional helicity-changing \textit{one-magnon} Brillouin light scattering, which apparently infringes the conservation law of the angular momentum. We show that the \textit{crystal angular momentum} intervenes to compensate the missing angular momentum in the latter scattering process.
\end{abstract}


\maketitle

For all the physical processes under the continuous rotational symmetry, the angular momentum is a good quantum number: it can only take quantized values, i.e., integers multiplied by the fundamental constant $\hbar$. Under such circumstances, the angular momentum is transferred from one agent to the other in such a way that the total amount is conserved. The angular momentum transfer occupies a central place in the modern development of spintronics. For example, spin pumping~\cite{JS1987}, spin transfer torque~\cite{RS2008}, and the spin Hall effect~\cite{SVWBJ2015} enable us to transfer angular momentum from electric currents to magnetization and vice versa. 

Angular momentum transfer occurs not only between spin-polarized electric current and magnetization, but also between polarized light and magnetization.  With magneto-optical effects such as the Faraday effect and the Cotton-Mouton effect, the coupling between polarized light and magnetization can be realized. Since at optical frequencies the magnetic dipole interaction ceases to play any role, the \textit{electric dipole moment} associated with the magnetization instead dictates the magneto-optical effects~\cite{LL8,Pershan}, which microscopically arise due to the (generally weak) spin-orbit coupling.

To investigate a further possibility of the magneto-optical effect in manipulating magnetization dynamics, let us turn our attention to the \textit{electric quadrupole moment}. For ferromagnetic (and ferrimagnetic) materials there has been much less interest in the electric quadrupole moment~\cite{KKR2010}, which would manifest itself in a process of a helicity-changing Brillouin light scattering in the \textit{Faraday geometry} (namely, light propagates parallel to the external magnetic field) as we discuss here. This is in stark contrast to the familiar one-magnon Brillouin light scattering in the \textit{Voigt geometry} (namely, the light propagates perpendicular to the external magnetic field), around which the emergent field of \textit{cavity optomagnonics}~\cite{Osada2016,Zhang2016,Haigh2016,Kusminskiy2016,Sharma2017,Osada2018,Haigh2018,Osada2018_2} is revolving. Nevertheless, for antiferromagnetic materials it is well known that the Brillouin light scattering by the quadrupole moment associated with two-magnon excitations is large when the two magnons involved in the scattering process originate in the modes with large and opposite wave-numbers~\cite{FPCG1966,FPL1967,Moriya1967,Moriya1968,FL1968}. In atomic physics, quadrupole moments of collective spin states have been widely studied~\cite{Polzik1999,Oberthaler2011,Klempt2011,Chapman2012} in connection with spin squeezing~\cite{KU1993}.

In this Letter, with a ferromagnetic spherical crystal, we experimentally explore the Brillouin light scattering in Faraday geometry using polarization-sensitive optical heterodyne measurements~\cite{Osada2016,Osada2018}. It is revealed that two-magnon excitations induce electric quadrupole moments, which give rise to the helicity-changing Brillouin light scattering. Besides, we find an unconventional helicity-changing Brillouin light scattering which only involves one-magnon excitations. For the latter case the conservation of the angular momentum is upheld only when the \textit{crystal angular momentum}~\cite{SB1968,Bloembergen1980,Nienhuis2002,Higuchi2011,Konishi2014} is taken into account. The possible relevance to the elusive rotational Doppler effect~\cite{SB1968} in the context of magnon-induced Brillouin light scattering is also discussed.

\begin{figure}[t]
\includegraphics[width=8.6cm,angle=0]{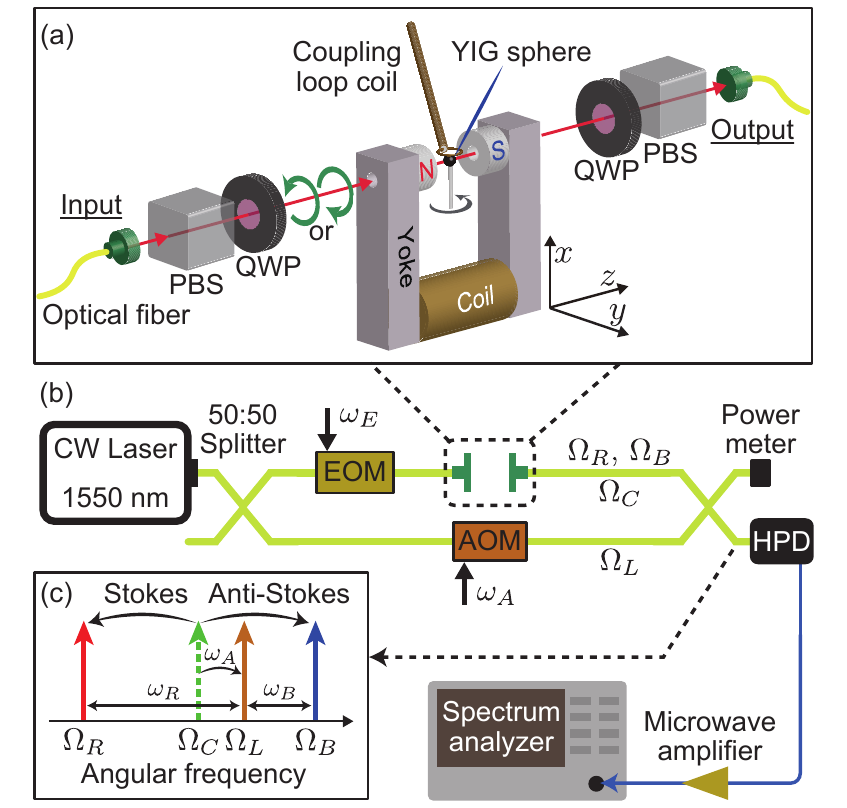}
\caption{(a)~A spherical crystal (0.5 mm in diameter) of yttrium iron garnet is placed in the gap of a magnetic circuit which consists of a pair of cylindrical permanent magnets, a coil, and a yoke. A coupling loop coil above the YIG sphere is used to excite magnons. By a set of a quarter-wave plate and a polarization beam splitter either left or right circularly polarized light is chosen for the input and the output. (b)~ Light from a cw laser is separated into two paths by an optical fiber splitter. An electro-optic modulator (EOM) in the upper path is used to calibrate signals and an acousto-optic modulator in the lower path is used to generate a local oscillator. The signal and the LO are combined and the resultant signal is sent to a power meter and a high-speed photo detector followed by a spectrum analyzer after a microwave amplifier. (c)~Schematic representation of the relevant frequencies. The carrier light at $\Omega_{C}$ is scattered into the sidebands at $\Omega_{R}$ and $\Omega_{B}$. The beat signals  appear at $\omega_{R}$ and $\omega_{B}$ with respect to the LO at $\Omega_{L}$.
}
\label{plot:setup}
\end{figure}

The experimental setup is schematically shown in Figs.~\ref{plot:setup}(a) and \ref{plot:setup}(b). A spherical crystal (0.5-mm in diameter) of yttrium iron garnet (YIG) is attached to an alumina rod oriented along the crystal axis $\langle110\rangle$ and placed at the center of the gap of a magnetic circuit as shown in Fig.~\ref{plot:setup}(a). The YIG sphere can be rotated about the $\langle110\rangle$ crystal axis, which allows us to apply a static magnetic field along either $\langle100\rangle$, $\langle111\rangle$, or any orientation in the (110) plane. The magnetic field, created by the magnetic circuit, around $130 \,\mathrm{kA/m}$ saturates the magnetization of the YIG sphere along the $z$-axis and can be varied. A coupling loop coil above the YIG sphere generates an oscillating magnetic field perpendicular to the saturated magnetization $M_{z}$ to excite magnons in the uniformly oscillating magnetostatic mode (Kittel mode) giving rise to the time-varying transverse magnetizations $M_{x}(t)$ and $M_{y}(t)$. 

\begin{figure}[t]
\includegraphics[width=8.6cm,angle=0]{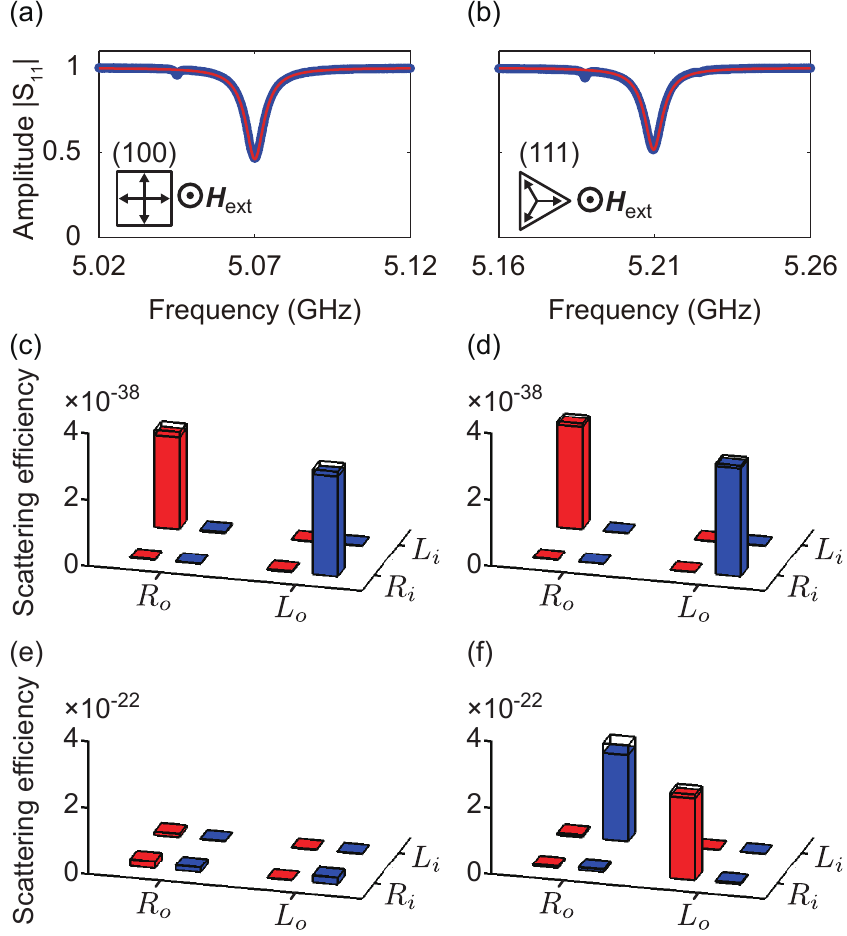}
\caption{(a) and (b): Microwave reflection spectrum for the Kittel mode under the external magnetic fields (a) $\bm{H}_{\mathrm{ext}}\parallel\langle100\rangle$ and (b) $\bm{H}_{\mathrm{ext}}\parallel\langle111\rangle$. The blue points show the measured reflection amplitude whereas the red curve shows the Lorentzian fitting. (c)-(f): Scattering efficiencies of the Stokes sideband (red bars) and the anti-Stokes sideband (blue bars) by two magnons ((c),(d)) and one magnon ((e),(f)) for four distinct polarization sets under $\bm{H}_{\mathrm{ext}}\parallel\langle100\rangle$ ((c),(e)) and $\bm{H}_{\mathrm{ext}}\parallel\langle111\rangle$ ((d),(f)). The height of the color bar shows the mean scattering efficiency and the difference between the top of the black wire frame and the bar represents a standard deviation estimated from measurements repeated six times. ${R}_{i}$ (${L}_{i}$) represents the right-circular (left-circular) polarization for the input field, while ${R}_{o}$ (${L}_{o}$) represents right-circular (left-circular) polarization for the output field.
}
\label{plot:2magnons}
\end{figure}

Figures~\ref{plot:2magnons}(a) and \ref{plot:2magnons}(b) show the microwave reflection spectra indicating the ferromagnetic resonances for the external magnetic field $\bm{H}_\mathrm{ext}$ being parallel to the $\langle100\rangle $ and $\langle111\rangle$ axes, respectively. From the fitting we obtain the resonance frequency of the Kittel mode $\omega_{K}/2\pi=5.07\,\mathrm{GHz}$ for $\bm{H}_{\mathrm{ext}}\parallel \langle100\rangle$ and $\omega_{K}/2\pi=5.21\,\mathrm{GHz}$ for $\bm{H}_{\mathrm{ext}}\parallel \langle111\rangle$.
Depending on the direction $\bm{H}_{\mathrm{ext}}$ with respect to the crystal axis, the magnon resonance angular frequency $\omega_{K}$ varies due to the magnetocrystalline anisotropy~\cite{GM,Stancil}, which is used to determine the crystal axis as described in Appendix~\cite{SM}.

We now explore the Brillouin light scattering under the condition in which the Kittel mode is continuously driven at the resonance. As shown in Fig.~\ref{plot:setup}(b) a cw laser light with a wavelength of 1550 nm (the angular frequency of $\Omega_{C}$) is split into two paths by an optical fiber splitter. The light in the lower path acts as a local oscillator (LO) whose frequency is shifted by $\omega_{A}/2\pi=80 \,\mathrm{MHz}$ from $\Omega_{C}$ by an acousto-optic modulator (AOM). In the upper path, the laser light is sent through the YIG sphere along the $z$-axis. In this Faraday geometry, the \textit{discrete} rotational symmetry is assured along the $z$-axis: for the case of $\bm{H}_{\mathrm{ext}}\parallel \langle100\rangle$ it is fourfold symmetry, and for the case of $\bm{H}_{\mathrm{ext}}\parallel \langle111\rangle$ it is threefold symmetry. By a pair of a quarter-wave plate (QWP) and a polarization beam splitter (PBS) before and after the YIG sphere as shown in Fig.~\ref{plot:setup}(a), either the left or right circularly-polarized light can be selected as the input and the output. The scattered light from the upper path interferes with the LO light from the lower path after the second optical fiber splitter so that the resultant beat signals originating from the Stokes scattering (red sideband) and the anti-Stokes scattering (blue sideband) appear at different angular frequencies, $\omega_{R}$ and $\omega_{B}$, respectively, as schematically shown in Fig.~\ref{plot:setup}(c). These beat signals are detected by a high-speed photodetector (HPD) and then amplified and analyzed by a spectrum analyzer. By using this setup we can investigate the selection rule both in the \textit{helicity-conserving} and the \textit{helicity-changing} Brillouin light scattering. 

Figure~\ref{plot:2magnons}(c) shows the observed two-magnon scattering efficiencies for the case of $\bm{H}_{\mathrm{ext}}\parallel \langle100\rangle$. The scattering efficiencies are deduced from the signal at the angular frequency of $\omega_{R}=2\omega_{K}+\omega_{A}$ for the two-magnon Stokes sideband and that at $\omega_{B}=2\omega_{K}-\omega_{A}$ for the two-magnon anti-Stokes sideband [see Appendix~\cite{SM} for a part of the raw data used to deduce the scattering efficiencies. The calibration scheme is also provided in Appendix~\cite{SM}. The same comment is applied to other scattering efficiencies shown in Figs~\ref{plot:2magnons}(d), \ref{plot:2magnons}(e), and \ref{plot:2magnons}(f)]. The significant helicity-changing two-magnon Stokes sideband appears when the input (output) polarization is left (right) circular (${L}_{i} \rightarrow {R}_{o}$ configuration), while the significant helicity-changing two-magnon anti-Stokes sideband appears when the input (output) polarization is right (left) circular (${R}_{i} \rightarrow {L}_{o}$ configuration). The fact that there is no signal when the microwave drive angular frequency is detuned from the resonance of the Kittel mode by $\delta \omega$ ($\delta \omega > \gamma_{t}$, where $\gamma_{t} \sim 2\pi \times 8\,\mathrm{MHz}$ is the linewidth of the Kittel mode) ensures the absence of the spurious drive signal coupled directly into the HPD.

As described in Appendix~\cite{SM}, the scattering efficiency for the helicity-changing two-magnon Stokes sideband at $\omega_{R}$ is proportional to the square of the electric quadrupole moment, $-\alpha M_{-}(t)^{2}$, in the ${L}_{i} \rightarrow {R}_{o}$ configuration, where $M_{-}(t)=M_{x}(t)-iM_{y}(t)$ is the transverse magnetization of the Kittel mode and $\alpha = \frac{G_{11}}{4} -\frac{G_{12}}{4}+ \frac{G_{44}}{2}$ with $G_{11}$, $G_{12}$, and $G_{44}$ are three parameters that specify the dielectric tensor of the cubic crystal (here, YIG). The scattering efficiency for the helicity-changing two-magnon anti-Stokes sideband at $\omega_{B}$ can be similarly explained with the electric quadrupole moment, $-\alpha M_{+}(t)^{2}$, with $M_{+}(t)=M_{x}(t)+iM_{y}(t)$. Note that in the quantum mechanical picture $M_{-}(t)^{2}$ ($M_{+}(t)^{2}$) corresponds to an operator which creates (annihilates) \textit{a pair of magnons}~\cite{SM}. Thus the helicity-changing two-magnon Brillouin light scattering can be viewed as a four-wave mixing process involving two magnons and two photons, which has been largely neglected. Note here that the Kittel mode with zero wave-number is diametrically opposed to the modes with large wave-numbers by which the four-wave mixing process involving two magnons and two photons has previously been observed with antiferromagnetic materials~\cite{FPCG1966,FPL1967,Moriya1967,Moriya1968,FL1968}.

\begin{figure}[t]
\includegraphics[width=8.6cm,angle=0]{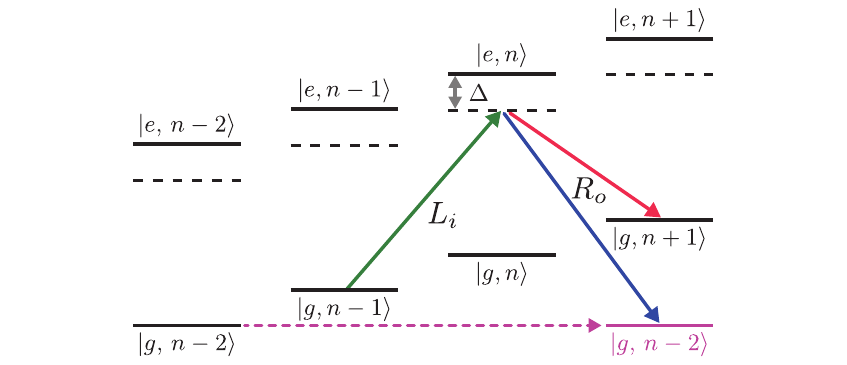}
\caption{Energy-level diagrams relevant to the Brillouin scattering.
The states are labeled by the electronic ground and excited states $|g\rangle$ and $|e\rangle$, respectively, with the number of magnons as $|n\rangle$. The green arrows represent the input carrier with the angular frequency of $\Omega_{C}$ and the red and blue arrows represent the Stokes and anti-Stokes sidebands with $\Omega_{R}$ and $\Omega_{B}$, respectively, for the ${L}_{i} \rightarrow {R}_{o}$ configuration. $\Delta$ denotes the frequency detuning between the light and the $|g\rangle \leftrightarrow|e\rangle$ transition. The horizontal dashed arrow connects the identical states due to the ambiguity emerged from the crystal angular momentum 3$\hbar$.
}
\label{plot:diagrams}
\end{figure}

To gain further insight, we now turn our attention to the case in which $\bm{H}_{\mathrm{ext}}\parallel \langle111\rangle$. Figure~\ref{plot:2magnons}(d) shows the observed helicity-changing and the helicity-conserving two-magnon scattering efficiencies for the case of $\bm{H}_{\mathrm{ext}}\parallel \langle111\rangle$. As in the case of $\bm{H}_{\mathrm{ext}}\parallel \langle100\rangle$ the significant Stokes sideband again appears in the ${L_{i}} \rightarrow {R_{o}}$ configuration while the significant anti-Stokes sideband appears in the ${R}_{i} \rightarrow {L}_{o}$ configuration. These sideband generation efficiencies agree with the ones theoretically predicted~\cite{SM}, which are proportional to the squares of the respective electric quadrupole moments, $-\beta M_{-}(t)^{2}$ and $-\beta M_{+}(t)^{2}$, with $\beta=\frac{G11}{6}-\frac{G_{12}}{6}+\frac{2G_{44}}{3}$, respectively. To see the situation schematically, Fig.~\ref{plot:diagrams} shows the energy-level diagrams relevant to the Brillouin scattering. The states $|g\rangle$ and $|e\rangle$ describe the electronic ground and excited states relevant to the dominant optical transition. Under the static magnetic field those states split to form a ladder depending on the magnon number, which is denoted by $|n\rangle$. Here, the helicity-changing two-magnon Stokes sideband in the ${L}_{i} \rightarrow {R}_{o}$ configuration observed in Fig.~\ref{plot:2magnons}(d), for instance, corresponds to the transition that connects $|g, n-1\rangle$ and $|g, n+1\rangle$ in Fig.~\ref{plot:diagrams}. In this transition, the angular momentum gained by light is $\Delta J_{p}=2\hbar$, while the same amount of angular momentum is lost from the sphere (i.e., $\Delta J_{m}=-2\hbar$) by creating two magnons (increasing magnon reduces the angular momentum of the sphere). Here, the total angular momenta are conserved among relevant two photons and two magnons and $\Delta J_{p} + \Delta J_{m} = 0$.


In the case of $\bm{H}_{\mathrm{ext}}\parallel \langle111\rangle$ a unusual situation appears when the scattering involves two photons and \textit{one} magnon, where the conservation of angular momentum is seemingly broken. As in the case of the two-magnon Brillouin light scattering, we obtain the one-magnon scattering efficiencies deduced from the signal observed at the angular frequency of $\omega_{R}=\omega_{K}+\omega_{A}$ for the Stokes sideband and that at $\omega_{B}=\omega_{K}-\omega_{A}$ for the anti-Stokes sideband, which are shown in Figs.~\ref{plot:2magnons}(e) and \ref{plot:2magnons}(f). In the case of $\bm{H}_{\mathrm{ext}}\parallel\langle100\rangle$, there is indeed no noticeable scattering as shown in Fig.~\ref{plot:2magnons}(e). In the case of $\bm{H}_{\mathrm{ext}}\parallel\langle111\rangle$, however, the significant helicity-changing Stokes sideband appears in the ${R}_{i} \rightarrow {L}_{o}$ configuration, and the significant helicity-changing anti-Stokes sideband appears in the ${L}_{i} \rightarrow {R}_{o}$ configuration as shown in Fig.~\ref{plot:2magnons}(f). This helicity-changing one-magnon anti-Stokes sideband generation in the ${L}_{i} \rightarrow {R}_{o}$ configuration, for instance, corresponds to the transition that connects $|g, n-1\rangle$ and $|g,n-2\rangle$ in Fig.~\ref{plot:diagrams}, where the angular momentum \textit{gained} in the sphere by annihilating one magnon is $\Delta J_{m}=\hbar$ even though the light also \textit{gains} the angular momentum by $\Delta J_{p}=2\hbar$.

The key to save the conservation of angular momentum is the \textit{crystal angular momentum} associated with the threefold symmetry possessed by the crystal along $\bm{H}_{\mathrm{ext}}\parallel\langle111\rangle$. Unlike isolated atoms, liquids, or amorphous solids, crystals do not have the \textit{continuous} rotational symmetry and thus the angular momentum is not a good quantum number. Angular momentum transfer processes taking place in the crystals are then determined up to the crystal angular momentum. This situation is analogous to the one that the linear momentum of an electron in the crystal has an ambiguity of $\hbar$ times reciprocal lattice vectors. In our particular example of the cubic crystal with $\bm{H}_{\mathrm{ext}}\parallel\langle111\rangle$, the crystal angular momentum is an integer multiple of 3$\hbar$. With this in mind, let us revisit the transition that connects $|g, n-1\rangle$ and $|g,n-2\rangle$ in Fig.~\ref{plot:diagrams}. The total angular momentum 3$\hbar$ gained by the sphere and the light can indeed be identified to be zero because of the ambiguity emerged from the crystal angular momentum 3$\hbar$ as indicated in Fig.~\ref{plot:diagrams}. 

These processes can thus be understood as a result of rotational analog of the umklapp process due to the crystal angular momentum. Note that in the standard group-theoretic analysis of selection rules in an inelastic scattering, everything is boiled down to the analysis of the excitation of the scatterer \textit{as a whole} in terms of the irreducible representations of its symmetry group~\cite{Tinkham,HL}. Thus, the origin of the angular momentum of the excitation (either coming from the one of magnetization or that of crystal, in our particular example) is usually not questioned. The importance of the crystal angular momentum has been argued in connection with the second harmonic generation~\cite{SB1968,Bloembergen1980,Konishi2014}, parametric down-conversion~\cite{Nienhuis2002}, and the Raman scattering by magnons with THz eigenfrequency in an antiferromagnetic material~\cite{Higuchi2011}. 

The fully continuous rotational symmetry for the Brillouin light scattering processes can be recovered if the rotational degree of freedom for the crystalline sphere as a whole is liberated. This can be done by considering the sphere as a freely rotating rigid body and introducing the azimuthal angle $\phi$ for the sphere along $\bm{H}_{\mathrm{ext}}\parallel\langle111\rangle$~\cite{SM}. The scattering efficiency of this sideband generation process now depends on $\phi$ and is proportional to the square of the electric quadrupole moment, $\xi M_{+}(t)$, with $\xi=\frac{\sqrt{2}}{3} e^{-3i \phi} g M_{z}$ as described in Appendix~\cite{SM}. Here $g=G_{11}-G_{12}-2G_{44 }$, which is zero when the material is isotropic. The phase factor $e^{-3i \phi}$ in $\xi$ can be considered as the spherical harmonics $Y_{3}^{-3}$ denoting the rotation of the spherical crystal. By writing the matrix element of the one-magnon transition amplitude from $|g, n-1\rangle$ to $|g,n-2\rangle$ in Fig.~\ref{plot:diagrams} as $\mathcal{U}_{o}$, the angular momentum of the sphere acquired in the course of the transition can be given by $\Delta J_{c}=\mathrm{Tr} \left[ \hat{\rho}(\mathcal{U}_{o}^{\dagger}\,\hat{L}_{z}\,\mathcal{U}_{o}-\hat{L}_{z}) \right]=-3\hbar$~\cite{SB1968,SM}, where $\hat{\rho}$ is the density matrix for the rotational state of the sphere and $\hat{L}_{z}$ is the $z$ component of the angular momentum operator for the sphere, whose Euler-angle representation reads $-i\hbar \frac{\partial}{\partial \phi}$. Thus, the excess angular momentum $\Delta J_{m}+\Delta J_{p} = 3 \hbar$ seen in the transition from $|g, n-1\rangle$ to $|g,n-2\rangle$ in Fig.~\ref{plot:diagrams} is indeed retrieved as the rotation of the sphere, leading to $\Delta J_{m}+\Delta J_{p}+\Delta J_{c}=0$. On the other hand, the matrix element of the two-magnon transition amplitude from $|g, n-1\rangle$ to $|g,n+1\rangle$ in Fig.~\ref{plot:diagrams} does not depend on $\phi$, meaning that there is no rotation of the sphere in this transition. The similar conclusion holds for other angular momentum transfers. We emphasize that the threefold discrete rotational symmetry of crystal is engraved deeply even when the full continuous rotational symmetry is resumed by liberating the rotational degree of freedom of the crystal. 

The nontrivial phase factor $e^{-3i \phi}$ in $\xi$ would give rise to an additional observable consequence: when the sphere is rotating along $\bm{H}_{\mathrm{ext}}\parallel\langle111\rangle$ at the angular velocity of $\omega_{L}$ the resultant sideband as a result of the one-magnon transition would experience the \textit{rotational Doppler shift} by $\Delta \omega_{K} = -3 \times \omega_{L}$. Here, the factor $-3$ stems from $e^{-3i \phi}$ and is basically what Simon and Bloembergen have predicted as early as in 1968 in the context of second harmonic generation~\cite{SB1968}. The predicted rotational Doppler shift has, to the best of our knowledge, not yet been observed. We envision that observing this shift in the magnon-induced Brillouin light scattering is feasible once a sphere is enforced to rotate uniformly or levitated to set free the rotation. We note that the levitation of a micron-scale ferromagnetic particles has been recently demonstrated~\cite{Budker2019,Hetet2019}.

In our experiment, the sphere is rigidly fixed on the optical table and the resultant moment of inertia is enormous. Since the rotational kinetic energy of the sphere acquired by the torque associated with the scattering is thus negligibly small, the intervention of the crystal angular momentum in the Brillouin light scattering would not affect the phase relationship between the input light and the scattered output light. As shown in Appendix~\cite{SM}, a set of measurements reveals that the phase relationship indeed remains unimpaired.

In summary, we demonstrated the presence of helicity-changing two-magnon scattering as well as the helicity-changing crystal-angular-momentum-assisted one-magnon scattering. We anticipate that the former process is ubiquitous in any ferro- and ferrimagnetic insulating materials supporting long wavelength magnetostatic modes. The latter process, however, only occurs in such materials with crystalline structure having threefold symmetry along the external magnetic field $\vector{H}_{\mathrm{ext}}$ in the Faraday geometry. 

We would like to thank Y.~Tabuchi, S.~Kono, A.~Okada, A.~Osada, G.~E.~W.~Bauer, K.~Sato, T.~Satoh,  E.~Saitoh, S.~Daimon, A.~Hatakeyama, H.~Watanabe, A.~Nunnenkamp, D.~Malz, A.~Ramsay, and J.~Haigh for useful discussion. This work is partly supported by KAKENHI (Grant no.~26220601) and JST-ERATO project (Grant no.~JP-MJER1601).

\appendix
\renewcommand{\thefigure}{S\arabic{figure}} 
\setcounter{figure}{0}

\section{Theory}

Suppose that magnons in a magnetostatic mode are excited in a ferromagnetic crystal. The resultant time-varying magnetization vector is represented by 
\begin{equation}
\vector{M}(t) = \left[
\begin{array}{l}
M_{x}(t) \\
M_{y}(t) \\
M_{z}
\end{array}
\right].
\end{equation}
Here the mean magnetization of the crystal is assumed to be along $z$-axis so that $M_{z}$ is large compared to the magnetizations in the plane perpendicular to $z$-axis and it is considered to be constant.

The light propagating through the crystal then experiences the time-varying magnetization $\vector{M}(t)$, which induces the polarization change of the light leading to the Brillouin scattering. In the off-resonant limit, the interaction Hamiltonian for the Brillouin light scattering can be given by
\begin{equation}
H_{\mathrm{OM}}(\tau) = \frac{\epsilon_0}{2} \int^{\tau+\delta}_{\tau} \mathcal{E} Ac'dt + \mbox{H.\,c.}, \label{eq:H}
\end{equation}
where $\epsilon_0$ is the permittivity of free space, $A$ is the cross section of the light beam, $\delta=\frac{l}{c'}$ is the interaction time with $c'$ being the speed of light in the material and $l$ being the interaction length, and $\mathrm{H.c.}$ represents the Hermitian conjugate of the first term~\cite{HM1967,Mabuchi2006,Polzik2010}. Here
\begin{equation}
\mathcal{E} = \left[(E')^{\ast}_{x}, (E')^{\ast}_{y}, (E')^{\ast}_{z}\right] \left[\begin{array}{ccc}
\epsilon_{xx} & \epsilon_{xy} & \epsilon_{xz} \\
\epsilon_{yx} & \epsilon_{yy} & \epsilon_{yz} \\
\epsilon_{zx} & \epsilon_{zy} & \epsilon_{zz} \\
\end{array} \right] \left[\begin{array}{c}
E_{x} \\
E_{y} \\
E_{z}
\end{array} \right] \label{eq:Hd}
\end{equation}
is the Hamiltonian density with $\epsilon_{ij}$ being the $ij$-th component of the second-rank dielectric tensor, $E_{i}$ and $(E')^{*}_{i}$ being the $i$-th components of the incident and scattered electric field, respectively. 

\subsection{Constraints on $\epsilon_{ij}$} 
Phenomenologically, we can understand the Brillouin light scattering by considering the tensor $\epsilon_{ij}$ in Eq.~(\ref{eq:H}) depending on the magnetization $\vector{M}(t)$~\cite{LL8,Pershan}. In powers of magnetization $\vector{M}=(M_{x},M_{y},M_{z})$, $\epsilon_{ij}$ can be expressed as
\begin{eqnarray}
\epsilon_{ij}(\vector{M})=\epsilon_{ij}^{0}+K_{ijk}M_{k}&+&G_{ijkl}M_{k}M_{l}+\cdots \label{eq:ep0}
\end{eqnarray}
where the repeated indices in each term are assumed to be summed up. Here, $\epsilon_{ij}^{0}$ is the permittivity independent of magnetization and is henceforth ignored. The next two terms involve the effect of the magnetization with the third-rank tensor $K_{ijk}$ and the fourth-rank tensor $G_{ijkl}$, which describe the strength of the coupling between the light and the magnetization and are endowed with the crystal symmetry~\cite{Moriya1967,Moriya1968}. Moreover, since we are dealing with the off-resonant limit of Brillouin light scattering, the photon number is preserved in the course of the scattering and thus the tensor $\epsilon_{ij}$ assumes hermiticity, i.e.,
\begin{equation}
\epsilon_{ji}^{*}(\vector{M}) = \epsilon_{ij}(\vector{M}). \label{eq:epH}
\end{equation}
This means that the real (imaginary) part of $\epsilon_{ji}$ is symmetric (antisymmetric). The Onsager relation further constrains the form of the tensor as 
\begin{equation}
\epsilon_{ij}(\vector{M}) = \epsilon_{ji}(-\vector{M}). \label{eq:epO}
\end{equation}
These constraints require the second (third) term in Eq.~(\ref{eq:ep0}) being antisymmetric (symmetric) for being odd (even) under the transformation $\vector{M} \rightarrow -\vector{M}$. 

With these requirements, $\epsilon_{ij}$ in Eq.~(\ref{eq:ep0}) up to the second order in the magnetizations can be reduced into
\begin{equation}
\epsilon_{ij}(\vector{M})= i \tilde{K}_{ijk}M_{k} + \bar{G}_{ijkl} \Sigma_{kl}, \label{eq:ep}
\end{equation}
where $\tilde{K}_{ijk}$ is the third-rank real tensor which is antisymmetric with respect to the indices $ij$, while $\bar{G}_{ijkl}$ is the fourth-rank real tensor which is symmetric with respect to the indices $ij$. Here, $\Sigma_{ij}$ are the symmetrized products of the two magnetization components, 
\begin{eqnarray}
\Sigma_{xx}(t) &=& M_x(t) M_x(t) \\
\Sigma_{yy}(t) &=& M_y(t) M_y(t) \\
\Sigma_{zz} &=& M_z M_z \\
\Sigma_{xy}(t) &=& 2M_x(t) M_y(t) \\
\Sigma_{yz}(t) &=& 2M_y(t) M_z \\
\Sigma_{xz}(t) &=& 2M_x(t) M_z. 
\end{eqnarray}
The first antisymmetric and imaginary term on the right-hand side of Eq.~(\ref{eq:ep}) represents the Faraday rotation of the plane of polarization while the second symmetric and real term represents the Cotton-Mouton effect.

The symmetrized products of the magnetization components $\Sigma_{ij}$ play a similar role to the symmetric strain tensor in the elasticity. A paramount difference is, however, the fact that, in the quantum mechanical interpretation, $\Sigma_{ij}$ involve two-magnon excitations (that is, $\Sigma_{xx}(t)$, $\Sigma_{yy}(t)$, and $\Sigma_{xy}(t)$, which do not contain $M_{z}$) while the symmetric strain tensor involves at most one-phonon excitation. This difference stems from the fact that the strain tensor in the elasticity contains the displacement vector $\vector{u}(t)$ only once.

The Cotton-Mouton term in the dielectric tensor represented by $\bar{G}_{ijkl}\Sigma_{kl}$ in Eq.~(\ref{eq:ep}) thus offers the Brillouin light scattering by two-magnon excitations. Hitherto, the Brillouin light scattering originating from the term associated with $\Sigma_{xx}(t)$, $\Sigma_{yy}(t)$, and $\Sigma_{xy}(t)$ are largely, if not completely, ignored for ferromagnetic crystals.  As for antiferromagetic crystals, the Raman scattering experiments have revealed the noticeable two-magnon excitations~\cite{FPCG1966,FPL1967,FL1968}. Here, simultaneous excitation of two magnons, which belong to the modes originating from different sub-lattices of the antiferromagnet, does not change the total magnetization. Such Brillouin light scattering by two-magnon excitations can be realized without enduring the spin-orbit interaction but exploiting much stronger exchange interaction, especially when these magnons come from the opposite edges of the Brillouin zone~\cite{Moriya1967,Moriya1968,FL1968}.

\subsection{Dielectric tensor for a cubic crystal}
Let us now consider the specific case with particular crystal symmetry, the cubic crystal. 

\subsubsection{Case I: $\bm{H}_{\mathrm{ext}}\parallel\langle100\rangle$}
For a cubic crystal (like YIG) with $\bm{H}_{\mathrm{ext}}\parallel\langle100\rangle$, the antisymmetric imaginary term $\tilde{K}_{ijk}M_{k}$ in Eq.~(\ref{eq:ep}) can be written in a matrix form with a parameter $f$ as
\begin{equation}
\begin{pmatrix}
0 & ifM_z & -ifM_y(t) \\
-ifM_z & 0 & ifM_x(t) \\
ifM_y(t) & -ifM_x(t) & 0 \\
\end{pmatrix}. \label{eq:ep2}
\end{equation}
As for the symmetric real term $\bar{G}_{ijkl} \Sigma_{kl}$ in Eq.~(\ref{eq:ep}), there are three independent parameters $G_{11}$, $G_{12}$, and $G_{44}$, leading to 
\begin{widetext}
\begin{equation}
\scalebox{1}{$\begin{pmatrix}
G_{11}\Sigma_{xx}(t)+G_{12}\Sigma_{yy}(t)+G_{12}\Sigma_{zz} & G_{44}\Sigma_{xy}(t) & G_{44}\Sigma_{xz}(t) \\
G_{44}\Sigma_{xy}(t) & G_{12}\Sigma_{xx}(t)+G_{11}\Sigma_{yy}(t)+G_{12}\Sigma_{zz} & G_{44}\Sigma_{yz}(t) \\
G_{44}\Sigma_{xz}(t) & G_{44}\Sigma_{yz}(t) & G_{12}\Sigma_{xx}(t)+G_{12}\Sigma_{yy}(t)+G_{11}\Sigma_{zz} \\
\end{pmatrix}$}. \label{eq:ep3}
\end{equation}
\end{widetext}

Let us see the relation between the dielectric tensor components $\epsilon_{ij}=\epsilon_{ij}^{'}+i \epsilon_{ij}^{''}$ and the magnetization $M_{i}$ in a slightly different way. The three imaginary components are trivially written as  
\begin{equation}
\begin{pmatrix}
{\epsilon}_{xy}^{''} \\
{\epsilon}_{yz}^{''} \\
{\epsilon}_{zx}^{''}
\end{pmatrix}
=if
\begin{pmatrix}
0 & 0 & 1 \\
1 & 0 & 0 \\
0 & 1 & 0 
\end{pmatrix}
\begin{pmatrix}
M_x(t) \\
M_y(t) \\
M_z 
\end{pmatrix}.
\label{eq:K}
\end{equation}
The six real components of the real symmetric tensor, Eq.~(\ref{eq:ep3}), can be represented as
\begin{equation}
\begin{pmatrix}
{\epsilon}_{xx}^{'} \\
{\epsilon}_{yy}^{'} \\
{\epsilon}_{zz}^{'} \\
{\epsilon}_{xy}^{'} \\
{\epsilon}_{yz}^{'} \\
{\epsilon}_{zx}^{'}
\end{pmatrix}
=
\begin{pmatrix}
      G_{11} & G_{12} & G_{12} & 0       & 0       & 0       \\
      G_{12} & G_{11} & G_{12} & 0       & 0       & 0       \\
      G_{12} & G_{12} & G_{11} & 0       & 0       & 0       \\
      0       &  0       & 0       & G_{44} & 0       & 0       \\
      0       &  0       & 0       & 0       & G_{44} & 0       \\
      0       &  0       & 0       & 0       & 0       & G_{44} 
\end{pmatrix}
\begin{pmatrix}
\Sigma_{xx}(t) \\
\Sigma_{yy}(t) \\
\Sigma_{zz} \\
\Sigma_{xy}(t) \\
\Sigma_{yz}(t) \\
\Sigma_{zx}(t) 
\end{pmatrix}.
\label{eq:G100}
\end{equation}
Notice that the form in Eq.~(\ref{eq:G100}) is analogous to the relation between the stress tensor and the strain tensor for a cubic crystal with the $6 \times 6$ matrix playing the role of the elastic stiffness matrix~\cite{LL7,Nye}. The analogy extends to the other cases with different crystal symmetries.

\subsubsection{Case II: $\bm{H}_{\mathrm{ext}}\parallel\langle111\rangle$}
Let us now consider a cubic crystal with $\bm{H}_{\mathrm{ext}}\parallel\langle111\rangle$. In this case, the fourfold rotational symmetry along the $z$-axis, held in the previous case, no longer exists. Instead, the rotational symmetry is reduced to threefold. 

The dielectric tensor $\epsilon_{ij}(\vector{M})$ in Eq.~(\ref{eq:ep}) is accordingly modified~\cite{Nye}. While the three imaginary components are unchanged from Eq.~(\ref{eq:K}), the six real components of the real symmetric tensor can now be represented as
\begin{widetext}
\begin{equation}
\begin{pmatrix}
{\epsilon}_{xx}^{'} \\
{\epsilon}_{yy}^{'} \\
{\epsilon}_{zz}^{'} \\
{\epsilon}_{xy}^{'} \\
{\epsilon}_{yz}^{'} \\
{\epsilon}_{zx}^{'}
\end{pmatrix}
=
\begin{pmatrix}
      G_{11} - \frac{g}{2} & G_{12} + \frac{g}{6} & G_{12} + \frac{g}{3} & 0       & -\frac{g}{3\sqrt{2}}       & 0       \\
      G_{12} + \frac{g}{6} & G_{11} - \frac{g}{2} & G_{12} + \frac{g}{3} & 0       & \frac{g}{3\sqrt{2}}       & 0       \\
      G_{12} + \frac{g}{3} & G_{12} + \frac{g}{3} & G_{11} - \frac{2 g}{3} & 0       & 0       & 0       \\
      0       &  0       & 0       & G_{44} + \frac{g}{3}  & 0       & \frac{g}{3\sqrt{2}}       \\
     -\frac{g}{3\sqrt{2}}       &  \frac{g}{3\sqrt{2}}       & 0       & 0       & G_{44} + \frac{g}{3} & 0       \\
      0       &  0       & 0       & \frac{g}{3\sqrt{2}}       & 0       & G_{44} + \frac{g}{6} 
\end{pmatrix}
\begin{pmatrix}
\Sigma_{xx}(t) \\
\Sigma_{yy}(t) \\
\Sigma_{zz} \\
\Sigma_{xy}(t) \\
\Sigma_{yz}(t) \\
\Sigma_{zx}(t) 
\end{pmatrix},
\label{eq:G111}
\end{equation}
\end{widetext}
where $g=G_{11}-G_{12}-2G_{44}$. Note that for the non-crystalline isotropic materials $g=0$ since the three parameters $G_{11}$, $G_{12}$, and $G_{44}$, are now written with just two parameters, $\lambda_{L}$ and $\mu_{L}$, by $G_{11}=\lambda_{L}+2\mu_{L}$, $G_{12}=\lambda_{L}$, and $G_{44}=\mu_{L}$. In the theory of elasticity $\lambda_{L}$ and $\mu_{L}$ are called the \textit{Lam\'{e} coefficients}~\cite{LL7}. The difference between Eqs.~(\ref{eq:G100}) and (\ref{eq:G111}) is due to the non-zero $g$, which can be considered as a consequence of rotational symmetry breaking associated with crystalline materials. 

\subsection{Dielectric tensor in the spherical basis}
In investigating the angular momentum transfer in the process of the Brillouin light scattering it is instrumental to invoke the spherical basis, which is widely used in atomic physics~\cite{BKD}, even though the crystals do not hold the full continuous rotation symmetry. The electric field in the Cartesian basis is given by
\begin{equation}
\vector{E} = E_{x} \vector{e}_{x}+E_{y} \vector{e}_{y} + E_{z} \vector{e}_{z},
\end{equation}
while in the spherical basis it is 
\begin{equation}
\vector{E} = E_{R} \vector{e}_{R}^{*}+E_{0} \vector{e}_{0}^{*} + E_{L} \vector{e}_{L}^{*},
\end{equation}
where the spherical basis $\left\{ \vector{e}_{R}, \vector{e}_{0}, \vector{e}_{L}\right\}$ is related to the Cartesian basis $\left\{ \vector{e}_{x}, \vector{e}_{y}, \vector{e}_{z}\right\}$ as~\cite{BKD}
\begin{equation}
\left[ \begin{array}{c}
\vector{e}_{R} \\
\vector{e}_{0} \\
\vector{e}_{L} 
\end{array}
\right] = \left[ \begin{array}{c}
-\frac{1}{\sqrt{2}} \left( \vector{e}_{x} + i \vector{e}_{y} \right) \\ 
\vector{e}_{z} \\
\frac{1}{\sqrt{2}} \left( \vector{e}_{x} - i \vector{e}_{y} \right) 
\end{array} \right].
\end{equation}
Here, along the reference axis (here, $z$-axis) $E_{R}$ and $E_{L}$ can be interpreted as the right- and left-circularly polarized components of the electric field while $E_{0}$ is the longitudinal component. With the spherical basis, the Hamiltonian density Eq.~(\ref{eq:Hd}) can be rewritten as
\begin{equation}
\mathcal{E} = \left[(E')^{\ast}_{R}, (E')^{\ast}_{0}, (E')^{\ast}_{L}\right] \left[\begin{array}{ccc}
\epsilon_{RR} & \epsilon_{R0} & \epsilon_{RL} \\
\epsilon_{0R} & \epsilon_{00} & \epsilon_{0L} \\
\epsilon_{LR} & \epsilon_{L0} & \epsilon_{LL} \\
\end{array} \right] \left[\begin{array}{c}
E_{R} \\
E_{0} \\
E_{L}
\end{array} \right]. \label{eq:Hds}
\end{equation}

Now let us restrict our interest to the case where the input (output) beam in the Brillouin light scattering comes in (out) along the $z$-axis, that is, along $\bm{H}_{\mathrm{ext}}$. In this case, $E_{0},$ and $(E')^{\ast}_{0}$ in Eq.~(\ref{eq:Hds}) disappear within the plane wave approximation. It is then convenient to introduce the Stokes parameters
\begin{eqnarray}
S_{x} &=& \frac{1}{2} \left( (E')_{R}^{\ast} E_{L} + (E')_{L}^{\ast} E_{R} \right) \\
S_{y} &=& -\frac{i}{2} \left( (E')_{R}^{\ast} E_{L} - (E')_{L}^{\ast} E_{R} \right) \\
S_{z} &=& \frac{1}{2} \left( (E')_{R}^{\ast} E_{R} - (E')_{L}^{\ast} E_{L} \right),
\end{eqnarray}
and the total power
\begin{equation}
S_{0}= (E')_{R}^{\ast} E_{R} + (E')_{L}^{\ast} E_{L},
\end{equation}
which are formed by $E_{R}$, $E_{L}$, $(E')^{\ast}_{R}$, and $(E')^{\ast}_{L}$. It is also useful to define
\begin{eqnarray}
S_{+} &=& S_{x}+iS_{y} \label{eq:s+}\\
S_{-} &=& S_{x}-iS_{y}. \label{eq:s-}
\end{eqnarray}
With these parameters denoting the combined polarization state of the input and output fields, the Hamiltonian density, Eq.~(\ref{eq:Hds}), can be expressed in the illuminating way~\cite{HM1967,Mabuchi2006,Polzik2010} as we shall see in the following.

\subsubsection{Case I: $\bm{H}_{\mathrm{ext}}\parallel\langle100\rangle$}
First, let us consider the case with $\bm{H}_{\mathrm{ext}}\parallel\langle100\rangle$. The Hamiltonian density can then be expressed as
\begin{equation}
\mathcal{E} = \mathcal{E}_{0} + \mathcal{E}_{1} + \mathcal{E}_{2}, \label{eq:Hds2}
\end{equation}
where
\begin{equation}
\mathcal{E}_{0}=\frac{1}{3} \left( G_{11}+2G_{12} \right) M^{2} S_{0} \label{eq:E0}
\end{equation}
is the contribution from the \textit{monopole} moment of the dielectric tensor, while  
\begin{equation}
\mathcal{E}_{1}= -f M_{z} S_{z} \label{eq:E1}
\end{equation}
and 
\begin{eqnarray}
\mathcal{E}_{2} &=& \frac{1}{6} \left( G_{11} -G_{12} \right) \left( M_{0}^{2} - 3M_{z}^{2} \right) S_{0} \nonumber \\
&\ & - \left[ \left(\frac{g}{4} +G_{44} \right)M_{-}^{2} + \frac{g}{4} M_{+}^{2} \right] S_{+}  \nonumber \\
&\ & - \left[ \left(\frac{g}{4} +G_{44} \right)M_{+}^{2} + \frac{g}{4} M_{-}^{2} \right] S_{-} \label{eq:E2}
\end{eqnarray}
are those from \textit{dipole} moment and the \textit{quadrupole} moment, respectively. Here $M^{2}=M_{x}^{2}+M_{y}^{2}+M_{z}^{2}$ is the total magnetization, $M_{+}=M_{x}+iM_{y}$ and $M_{-}=M_{x}-iM_{y}$ are the normal modes of the transverse magnetizations, $M_{x}$ and $M_{y}$. To prepare the quantum-mechanical analyses of scattering processes, let us here discuss the quantum-mechanical aspects of the normal modes $M_{+}$ and $M_{-}$. Quantum mechanically, $M_{+}$ ($M_{-}$) annihilates (creates) a magnon and increases (reduces) the magnetization. These normal modes read 
\begin{eqnarray}
M_{+}(t)=\hat{M}_{+}e^{-i \omega_{K}t}=-\gamma \hbar \frac{\sqrt{N}}{V} \hat{c} e^{-i \omega_{K}t}
\end{eqnarray}
and 
\begin{eqnarray}
M_{-}(t) = \hat{M}_{-}e^{i \omega_{K}t} = -\gamma \hbar \frac{\sqrt{N}}{V} \hat{c}^{\dagger} e^{i \omega_{K}t},
\end{eqnarray}
respectively, where $\hat{c}$ and $\hat{c}^{\dagger}$ are the annihilation and creation operators, $\gamma$ is the gyromagnetic ratio, $V$ is the volume of the sphere, and $N$ is the number of spins in the sphere~\cite{Kittel}. 

The monopole and the quadrupole terms involving $S_{0}$ in Eqs.~(\ref{eq:E0}) and (\ref{eq:E2}), respectively, do not change the polarization state of light; they merely introduce overall phase change that usually of no interest. The dipole term involving $S_{z}$ in Eq.~(\ref{eq:E1}) is responsible for the Faraday effect; the plane of polarization rotates in proportion to the magnetization along the $z$-axis, that is, $M_{z}$, which is static in our case. 

The quadrupole term involving $S_{+}= (E')_{R}^{\ast} E_{L}$ in Eq.~(\ref{eq:E2}) represents the helicity-changing Brillouin light scattering observable with the ${L}_{i} \rightarrow {R}_{o}$ configuration. There are two contributions from the quadrupole moment: one is from the term
\begin{equation}
-\alpha \hat{M}_{-}(t)^{2} = -\left(\frac{g}{4} +G_{44} \right)\hat{M}_{-}^{2}e^{i 2\omega_{K}t},
\end{equation}
and the other is from 
\begin{equation}
-\alpha'\hat{M}_{+}(t)^{2} = -\frac{g}{4} \hat{M}_{+}^{2}e^{-i 2\omega_{K}t},
\end{equation}
where $\alpha= \frac{G_{11}}{4}-\frac{G_{12}}{4}+\frac{G_{44}}{2}$ and $\alpha'=\frac{G_{11}}{4}-\frac{G_{12}}{4}-\frac{G_{44}}{2}$. The former appears as a helicity-changing two-magnon Stokes sideband at the angular frequency $2\omega_{K}$, whose scattering efficiency \textit{per magnon pair}, that is, $F_{s}/\left( F_{C} \times n^{2} \right)$, is proportional to 
\begin{eqnarray}
& & \frac{1}{n^{2}}\left[ \left(\frac{g}{4} +G_{44} \right) \langle n+1 | \hat{M}_{-}^{2} | n-1 \rangle \right]^{2} \nonumber \\ 
&=& \left[ \left(\frac{g}{4} +G_{44} \right) \right]^{2} \left(\frac{\sqrt{n(n+1)}}{n}\right)^{2} \left(\gamma \hbar \right)^{4} \frac{N^{2}}{V^{4}} \nonumber \\
&\sim& \left[ \left(\frac{g}{4} +G_{44} \right) \right]^{2} \left(\gamma \hbar \right)^{4} \frac{N^{2}}{V^{4}},
\end{eqnarray}
where $F_{C}$ is the carrier photon flux, $F_{s}$ is the scattered photon flux, and $n$ is the number of magnons excited in the sphere. On the other hand, the latter appears as a helicity-changing two-magnon anti-Stokes sideband at $-2\omega_{K}$, whose scattering efficiency \textit{per magnon pair} is proportional to 
\begin{eqnarray}
& & \frac{1}{n^{2}}\left[ \frac{g}{4} \langle n-1 |\hat{M}_{+}^{2} | n+1 \rangle \right]^{2} \nonumber \\
&=& \left[ \frac{g}{4} \right]^{2} \left(\frac{\sqrt{n(n+1)}}{n}\right)^{2} \left(\gamma \hbar \right)^{4} \frac{N^{2}}{V^{4}} \nonumber \\
&\sim& \left[ \frac{g}{4} \right]^{2} \left(\gamma \hbar \right)^{4} \frac{N^{2}}{V^{4}}.
\end{eqnarray}
The former scattering efficiency is about 50 times larger than the latter since the the value $\frac{g}{4}$ is evaluated to be about 7 times smaller than $\frac{g}{4} +G_{44}$ in the case of YIG with $gM_{s}^2=5.73\times10^{-5}$ and $G_{44}M_{s}^2=-1.14\times10^{-4}$, where $M_{s}=140$~kA/m is the saturation magnetization~\cite{Stancil}. Thus the latter scattering is too small to be seen experimentally because of the noise level (see Appendix~C). 

The quadrupole term involving $S_{-}= (E')_{L}^{\ast} E_{R}$ in Eq.~(\ref{eq:E2}) represents the helicity-changing Brillouin light scattering observable with the ${R}_{i} \rightarrow {L}_{o}$ configuration. Analysis of the scattering in this configuration follows the same line as above.

\subsubsection{Case II: $\bm{H}_{\mathrm{ext}}\parallel\langle111\rangle$}
Next, we shall consider the case with $\bm{H}_{\mathrm{ext}}\parallel\langle111\rangle$. The Hamiltonian density originates from the monopole moment of the dielectric tensor, $\mathcal{E}_{0}$, and that from the dipole moment, $\mathcal{E}_{1}$ are in fact unchanged from Eqs.~(\ref{eq:E0}) and (\ref{eq:E1}), respectively. On the other hand, we have
\begin{eqnarray}
\mathcal{E}_{2} &=& \frac{1}{6} \left( G_{11} -G_{12}- g \right) \left( M_{0}^{2} - 3M_{z}^{2} \right) S_{0} \nonumber \\
&\ & - \left[ \left(\frac{g}{6} +G_{44} \right)M_{-}^{2} - \frac{\sqrt{2}g}{3} M_{z}M_{+} \right] S_{+} \nonumber \\
&\ & - \left[ \left(\frac{g}{6} +G_{44} \right)M_{+}^{2} - \frac{\sqrt{2}g}{3} M_{z}M_{-} \right] S_{-} \label{eq:E2_2}
\end{eqnarray}
as the contribution from the quadrupole moment.

The quadrupole term involving $S_{+}= (E')_{R}^{\ast} E_{L}$ in Eq.~(\ref{eq:E2_2}) represents the helicity-changing Brillouin light scattering observable with the ${L}_{i} \rightarrow {R}_{o}$ configuration. There are two contributions from the quadrupole moment: one is from the term 
\begin{equation}
-\beta \hat{M}_{-}(t)^{2} = -\left(\frac{g}{6} +G_{44} \right)\hat{M}_{-}^{2}e^{i 2\omega_{\mathrm{K}}t}
\end{equation}
and the other is from 
\begin{equation}
\xi_{0} \hat{M}_{+}(t) = \frac{\sqrt{2}g}{3} M_{z}\hat{M}_{+}e^{-i \omega_{\mathrm{K}}t},
\end{equation}
where $\beta=\frac{G_{11}}{6}-\frac{G_{12}}{6}+\frac{2G_{44}}{3}$ and $\xi_{0}=\frac{\sqrt{2}g}{3} M_{z}$. The former appears as a helicity-changing two-magnon Stokes sideband at the angular frequency $2\omega_{\mathrm{K}}$, whose scattering efficiency \textit{per magnon pair}, that is, $F_{s}/\left( F_{C} \times n^{2} \right)$, is proportional to 
\begin{eqnarray}
& & \frac{1}{n^{2}}\left[ \left(\frac{g}{6} +G_{44} \right) \langle n+1 | \hat{M}_{-}^{2} | n-1 \rangle \right]^{2} \nonumber \\
&=& \left[ \left(\frac{g}{6} +G_{44} \right) \right]^{2} \left( \frac{\sqrt{n(n+1)}}{n} \right)^{2} \left(\gamma \hbar \right)^{4} \frac{N^{2}}{V^{4}} \nonumber \\
&\sim& \left[ \left(\frac{g}{6} +G_{44} \right) \right]^{2} \left(\gamma \hbar \right)^{4} \frac{N^{2}}{V^{4}}. \label{eq:seT}
\end{eqnarray}
On the other hand, the latter appears as a helicity-changing \textit{one-magnon} anti-Stokes sideband at $-\omega_{\mathrm{K}}$, whose scattering efficiency \textit{per magnon}, that is, $F_{s}/\left( F_{C} \times n \right)$, is proportional to
\begin{eqnarray}
& & \frac{1}{n}\left[ \frac{\sqrt{2}g}{3} M_{z} \langle n-2 | \hat{M}_{+} | n-1 \rangle \right]^{2} \nonumber \\
&=& \left[ \frac{\sqrt{2}g}{3} \right]^{2} \frac{n-1}{n} \left(\gamma \hbar \right)^{4} \frac{N^{3}}{4V^{4}} \nonumber \\
&\sim& \left[ \frac{\sqrt{2}g}{3} \right]^{2} \left(\gamma \hbar \right)^{4} \frac{N^{3}}{4 V^{4}}, \label{eq:seO}
\end{eqnarray}
where $M_{z}=\frac{1}{2}(-\gamma \hbar) \frac{N}{V}$. The ratio of the efficiency given by Eq.~(\ref{eq:seO}) to that given by Eq.~(\ref{eq:seT}), that is, 
\begin{equation}
\frac{\left[ \frac{\sqrt{2}g}{3} \right]^{2} \frac{N}{4}}{\left[ \left(\frac{g}{6} +G_{44} \right) \right]^{2}}
\end{equation}
is evaluated to be $2.3\times10^{16}$ in the case of YIG with $gM_{s}^2=5.73\times10^{-5}$, $G_{44}M_{s}^2= -1.14\times10^{-4}$, and $n_{s} = 2.11\times10^{28}\,\mathrm{m^{-3}}$, where $M_{s}=140$~kA/m is the saturation magnetization and $n_{s}$ is the density of the number of spins~\cite{Stancil}. This value is in reasonable agreement with the ratio $(0.84 \pm 0.10)\times10^{16}$ that is obtained from the carefully calibrated measurements [shown in Figs.~2(d) and (f) of the main text].

The quadrupole term involving $S_{-}= (E')_{L}^{\ast} E_{R}$ in Eq.~(\ref{eq:E2_2}) represents the helicity-changing Brillouin light scattering observable with the ${R}_{i} \rightarrow {L}_{o}$ configuration. Analysis of the scattering in this configuration follows the same line as above.

\subsection{Dielectric tensor with the azimuthal angle $\phi$}

Although crystals do not have the \textit{continuous} rotation symmetry, the symmetry can be resumed if the rotational degree of freedom for the crystal as a whole is liberated. This can be accomplished by making the azimuthal angle $\phi$, which specifies the crystal orientation with respect to the laboratory frame, an active variable. We shall now see how the Hamiltonian densities $\mathcal{E}_{2}$ for the case of $\bm{H}_{\mathrm{ext}}\parallel\langle100\rangle$ [Eq.~(\ref{eq:E2})] and that for  $\bm{H}_{\mathrm{ext}}\parallel\langle111\rangle$ [Eq.~(\ref{eq:E2_2})] would be modified by rotating the sphere by $\phi$ along $\bm{H}_{\mathrm{ext}}$.

\subsubsection{Case I: $\bm{H}_{\mathrm{ext}}\parallel\langle100\rangle$}
When the sphere is rotated by $\phi$ along $\bm{H}_{\mathrm{ext}}$, the Hamiltonian densities $\mathcal{E}_{2}$ given by Eq.~(\ref{eq:E2}) changes into
\begin{eqnarray}
&\ &\mathcal{E}_{2} = \frac{1}{6} \left( G_{11} -G_{12} \right) \left( M_{0}^{2} - 3M_{z}^{2} \right) S_{0} \nonumber \\
&\ & - \left[ \left(\frac{g}{4} +G_{44} \right)M_{-}^{2} + \frac{g}{4} e^{-4i \phi} M_{+}^{2} \right] S_{+}  \nonumber \\
&\ & - \left[ \left(\frac{g}{4} +G_{44} \right)M_{+}^{2} + \frac{g}{4} e^{4i \phi} M_{-}^{2} \right] S_{-}. \label{eq:E2p}
\end{eqnarray}
We see that only terms representing the transitions that apparently violate the conservation of angular momentum among photons and magnons acquire the extra phase factor $e^{\pm4i \phi}$. The phase factor $e^{-4i \phi}$ ($e^{4i \phi}$) comes in as a rotational degree of freedom of the crystal with fourfold symmetry. The resultant spherical harmonics $Y_{4}^{-4}$ ($Y_{4}^{4}$), with which $M_{+}^{2}$ ($M_{-}^{2}$) forms the quadrupole moment $Y_{2}^{-2}$ ($Y_{2}^{2}$), leads to the designated helicity-changing Brillouin light scattering. 

\subsubsection{Case II: $\bm{H}_{\mathrm{ext}}\parallel\langle111\rangle$}
When the sphere is rotated by $\phi$ along $\bm{H}_{\mathrm{ext}}$, the Hamiltonian densities $\mathcal{E}_{2}$ given by Eq.~(\ref{eq:E2_2}) changes into
\begin{eqnarray}
&\ & \mathcal{E}_{2} = \frac{1}{6} \left( G_{11} -G_{12}- g \right) \left( M_{0}^{2} - 3M_{z}^{2} \right) S_{0} \nonumber \\
&\ & - \left[ \left(\frac{g}{6} +G_{44} \right)M_{-}^{2} - \frac{\sqrt{2}g}{3}M_{z} e^{-3i \phi}M_{+} \right] S_{+} \nonumber \\
&\ & - \left[ \left(\frac{g}{6} +G_{44} \right)M_{+}^{2} - \frac{\sqrt{2}g}{3}M_{z} e^{3i \phi}M_{-} \right] S_{-}. \label{eq:E2_2p}
\end{eqnarray}
We see again that only terms representing the transitions that apparently violate the conservation of angular momentum among photons and magnons acquire the extra phase factor $e^{\pm3i \phi}$. The phase factor $e^{-3i \phi}$ ($e^{3i \phi}$) comes in as a rotational degree of freedom of the crystal with threefold symmetry. The resultant quadrupole moments read 
\begin{equation}
\xi \hat{M}_{+}(t) = \frac{\sqrt{2}g}{3} M_{z} e^{-3i \phi} \hat{M}_{+}e^{-i \omega_{\mathrm{K}}t}
\end{equation}
and 
\begin{equation}
\xi^{*} \hat{M}_{-}(t) = \frac{\sqrt{2}g}{3} M_{z} e^{3i \phi} \hat{M}_{-}e^{i \omega_{\mathrm{K}}t}
\end{equation}
with $\xi = \frac{\sqrt{2}g}{3} M_{z} e^{-3i \phi}$ and $\xi^{*} = \frac{\sqrt{2}g}{3} M_{z} e^{3i \phi}$, which induce the designated helicity-changing Brillouin light scattering.

Let us go a little bit further for this case. From Eqs.~(\ref{eq:H}) and (\ref{eq:E2_2p}) the helicity-changing transition amplitude associated with the Stokes parameter $S_{+}$ can be given as the first-order term of the perturbative expansion of the evolution operator~\cite{CDG}:
\begin{eqnarray}
\mathcal{U} &=& -\frac{1}{2i\hbar} \int_{t_{i}}^{t_{f}} d\tau \int_{\tau}^{\tau+\delta} dt \left[ \left( \frac{g}{6} + G_{44} \right) M_{-}^{2} \right. \nonumber \\
&\ & \left. -\frac{\sqrt{2} g}{3} M_{z} e^{-3i\phi} M_{+}  \right] Ac', \label{eq:U}
\end{eqnarray}
and that with $S_{-}$ by
\begin{eqnarray}
\mathcal{T} &=& -\frac{1}{2i\hbar} \int_{t_{i}}^{t_{f}} d\tau \int_{\tau}^{\tau+\delta} dt \left[ \left( \frac{g}{6} + G_{44} \right) M_{+}^{2} \right. \nonumber \\
&\ & \left. -\frac{\sqrt{2} g}{3} M_{z} e^{3i\phi} M_{-}  \right] Ac', \label{eq:T}
\end{eqnarray}
where, $t_{i}$ and $t_{f}$ are the initial and the final time of the transition. The terms in the amplitudes in Eqs.~(\ref{eq:U}) and (\ref{eq:T}) are classified by the phase factor with $\phi$, which determines whether the crystal angular momentum intervenes in the transition. For instance, the matrix element of transition amplitude $\mathcal{U}$ from $|g,n-1\rangle$ to $|g,n+1\rangle$ in Fig.~3 of the main text, that is,
\begin{eqnarray}
\mathcal{U}_{t} &=& - \frac{1}{2i\hbar} \int_{t_{i}}^{t_{f}} d\tau \int_{\tau}^{\tau+\delta} dt \left( \frac{g}{6} + G_{44} \right)\notag\\
&&\hspace{15mm}\times \langle n+1| M_{-}^{2}|n-1\rangle Ac', \label{eq:U1}
\end{eqnarray}
which denotes the two-magnon transition, has the trivial phase factor of unity. There is thus no intervention of the crystal angular momentum in the transition. On the other hand, the matrix element of transition amplitude $\mathcal{U}$ from $|g,n-1\rangle$ to $|g,n-2\rangle$, that is,
\begin{eqnarray}
\mathcal{U}_{o} &=& \frac{1}{2i\hbar} \int_{t_{i}}^{t_{f}} d\tau \int_{\tau}^{\tau+\delta} dt \frac{\sqrt{2}}{3} e^{-3i \phi} g M_{z}\notag\\
&&\hspace{15mm}\times \langle n-2|M_{+}|n-1\rangle Ac', \label{eq:U2}
\end{eqnarray}
which denotes the one-magnon transition, has the phase factor of $e^{-3i\phi}$. The crystal angular momentum serves, in this case, $-3\hbar$ in the transition. We can explicitly verify this by calculating the difference of the angular momentum of the sphere before and after the one-magnon transition:
\begin{eqnarray}
\Delta J_{c} &=&\mathrm{Tr}\left[ \hat{\rho}\left(\mathcal{U}_{o}^{\dagger}\,\hat{L}_{z}\,\mathcal{U}_{o}-\hat{L}_{z}\right) \right] \notag\\ 
&=&\sum_{n} p_{n} \langle\Psi_{n}| \left(\mathcal{U}_{o}^{\dagger}\,\hat{L}_{z}\,\mathcal{U}_{o}-\hat{L}_{z}\right)|\Psi_{n}\rangle\notag\\
&=&-3\hbar
\end{eqnarray}
where $\hat{\rho}$ is the density matrix for the rotational state of the sphere with $|\Psi_{n}\rangle$ denoting the eigenstates for the initial rotational state and $p_{n}$ denoting the probability of initially finding the state in $|\Psi_{n}\rangle$. Here, $\hat{L}_{z}$ is the $z$ component of the angular momentum operator for the sphere, whose Euler-angle representation reads $-i\hbar \frac{\partial}{\partial \phi}$. 

\begin{figure}[b]
\includegraphics[width=8.6cm,angle=0]{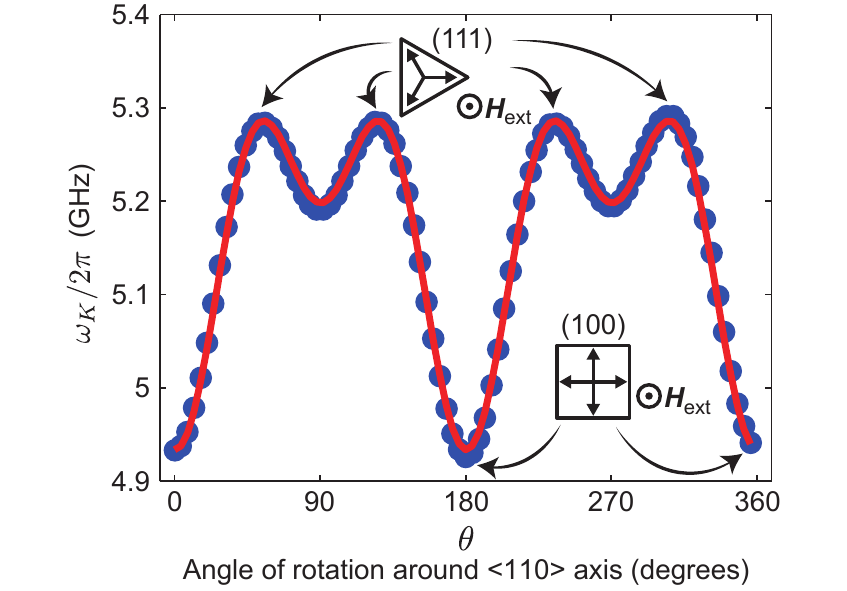}
\caption{Angular dependence of the resonance frequency of the Kittel mode, $\omega_{K}/2\pi$. Blue points show the measurement results and red curve shows the fitting based on Eq.~(\ref{eq:anisotropy}).
}
\label{plot:anisotropy}
\end{figure}

\section{Determination of crystal axis}
In this section, we explain how to determine the crystal axis of the YIG sphere. The blue points in Fig.~\ref{plot:anisotropy} show the measured angular dependence of the resonance frequency of  the Kittel mode $\omega_{K}/2\pi$.
This angular dependence can be written phenomenologically as~\cite{Healy}
\begin{eqnarray}
\frac{\omega_{K}}{2\pi} &=& \gamma \mu_0 \left( |H_{\mathrm{ext}}| \right. \nonumber \\
&&\hspace{-1.5mm} \left. + \frac{K_1}{\mu_0 M_{s}}\left(\frac{-16}{3}+\frac{5}{4}\cos 2\theta +\frac{15}{16}\cos 4\theta\right)\right) \label{eq:anisotropy}
\end{eqnarray}
where $\gamma$ is the electron gyromagnetic ratio, $\mu_0$ is the vacuum permeability, $|H|_{\mathrm{ext}}$ is the strength of the external magnetic field, $M_{s}$ is the saturation magnetization, $K_1$ is the constant with dimension of energy density representing magnetocrystalline anisotropy, and $\theta$ is the angle between the external magnetic field direction in the (110) plane and the $\langle100\rangle$ axis. The red curve shows the fitting result based on Eq.~(\ref{eq:anisotropy}), from which we obtain $|H_\mathrm{ext}|=130\,\mathrm{kA/m}$ and $K_{1}/M_{s}=-3.8\,\mathrm{mT}$. The value of $K_{1}/M_{s}$ is comparable to $-4.4\,\mathrm{mT}$ that is estimated from the result in the literature~\cite{Stancil}. The resonance frequency $\omega_{K}/2\pi$ takes the maximum (minimum) value for the case with $\bm{H}_{\mathrm{ext}}\parallel\langle111\rangle\,(\langle100\rangle)$. This measurement makes it possible to determine the crystal orientation with respect to $\bm{H}_{\mathrm{ext}}$.

\begin{figure*}[t]
\includegraphics[width=17.2cm,angle=0]{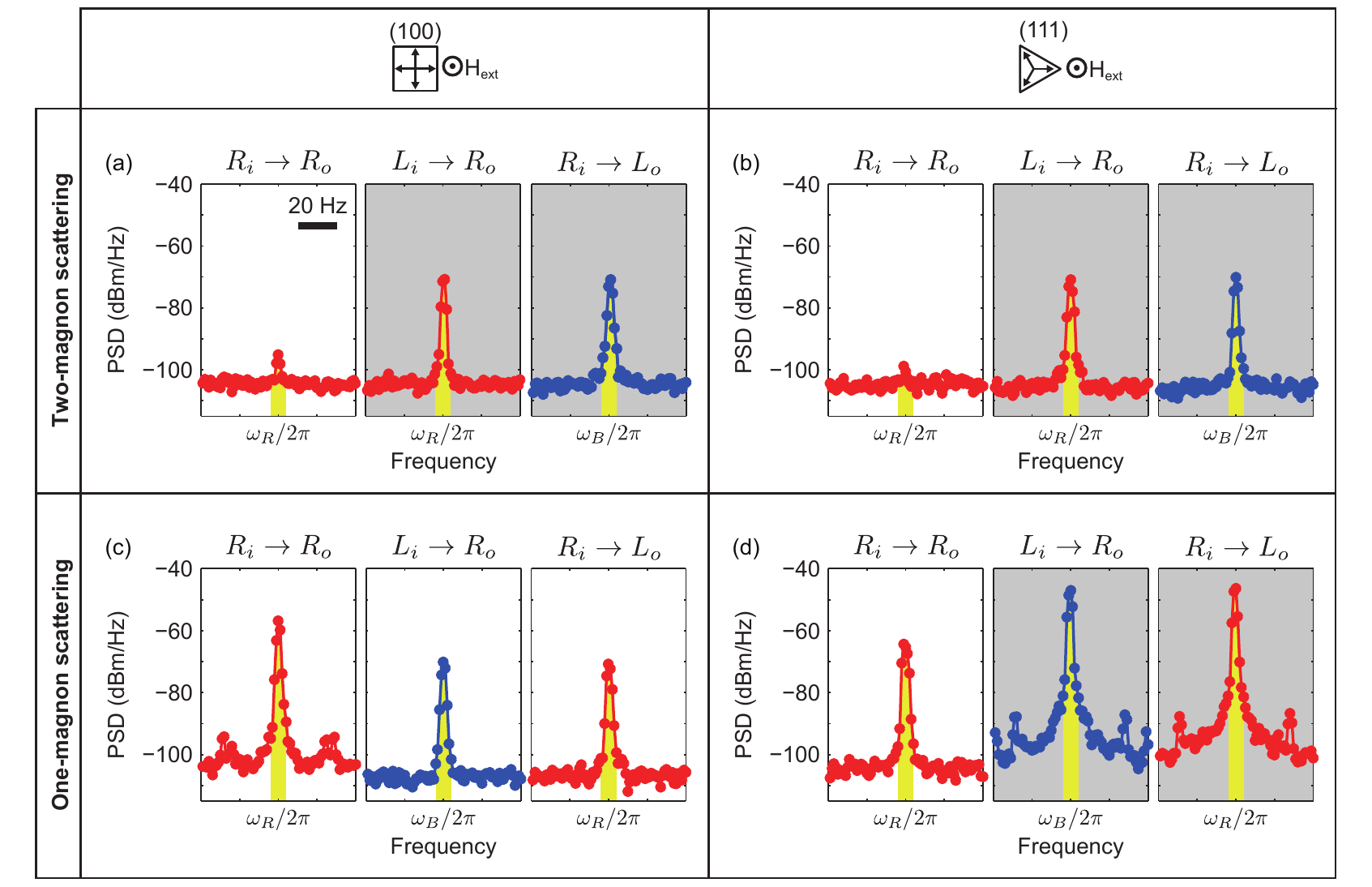}
\caption{Typical measured power spectra of the Stokes sidebands at angular frequency $\omega_{R}$ (red points) and the anti-Stokes sidebands at angular frequency $\omega_{B}$ (blue points). (a) Spectra for the two-magnon scattering under $\bm{H}_{\mathrm{ext}}\parallel\langle100\rangle$; (b) Spectra for the two-magnon scattering under $\bm{H}_{\mathrm{ext}}\parallel\langle111\rangle$; (c) Spectra for the one-magnon scattering under $\bm{H}_{\mathrm{ext}}\parallel\langle100\rangle$; (d) Spectra for the one-magnon scattering under $\bm{H}_{\mathrm{ext}}\parallel\langle111\rangle$. The figures with gray (white) background are used as signal (noise). Each signal (noise) power $P_{\mathrm{SA}}(\omega_{s})$ is obtained by integrating the power spectral density over the yellow region. Both the resolution bandwidth and the measurement point interval of the spectrum analyzer are set to be 1~Hz.}
\label{plot:PSD}
\end{figure*}

\section{Calibration of scattering efficiencies}
In this section, we explain how to deduce the scattering efficiencies shown in Fig.~2 from the power spectrum measured by the spectrum analyzer. The scattering efficiencies are calculated using the three values: the carrier photon flux $F_{C}$, the scattered photon flux $F_{s}$, and the number of magnons $n$ excited in the sphere. These values are deduced from experimental results. Here, the one-magnon scattering efficiencies are given by
\begin{equation}
\frac{F_{s}}{F_{C} \times n },
\end{equation}
while the two-magnon scattering efficiencies are given by 
\begin{equation}
\frac{F_{s}}{F_{C}\times n^2}.
\end{equation}
The method of evaluation of each value is described below. 

\subsection{Evaluation of $F_{C}$}
Since the scattering efficiency is very low, the power of the carrier light dominates the total optical power which can be measured by the power meter passing through the upper path shown in Fig.~1(b). 
The carrier photon flux $F_{C}$ can be simply deduced by dividing the light power $P_{C}$ by the photon energy $\hbar \Omega_{C}$, that is, 
\begin{equation}
F_{C}= \frac{P_{C}}{\hbar \Omega_{C}}.
\end{equation}

\subsection{Evaluation of $F_{s}$}
The scattered photon flux $F_{s}$ at the angular frequency $\Omega_{s}$ is related to the scattered power $P_{s}$ by 
\begin{equation}
F_{s}= \frac{P_{s}}{\hbar \Omega_{s}}. \label{eq:relation1}
\end{equation}
The problem here is that $P_{s}$ is too small to measure directly by power meter. The basic idea to deduce $P_{s}$ is to use a calibration tone at $\Omega_{t}$ with a calibrated power of $P_{t}$ which is generated by a calibrated EOM.
Then, between the known power of the optical calibration tone, $P_{t}(\Omega_{t})$, and the measured power of the beat signal, $P_{\mathrm{SA}}(\omega_{t})$, detected at the angular frequency of $\omega_{t}=|\Omega_{t}-\Omega_{L}|$ at the spectrum analyzer, we have the relation
\begin{equation}
P_{t}(\Omega_{t}) \times T(\Omega_{t} \rightarrow \omega_{t}) = P_{\mathrm{SA}}(\omega_{t}), \label{eq:relation0}
\end{equation}
where $T(\Omega_{t} \rightarrow \omega_{t})$ is the transfer function from the optical signal at $\Omega_{t}$ to the microwave signal at $\omega_{t}$, which collectively describes various effects such as the frequency variation of the sensitivity of the HPD and the gain of the microwave amplifier and the loss and the interference effect that the signal suffers in the coaxial cables. In this way, the transfer function $T(\Omega_{t} \rightarrow \omega_{t})$ is at first calibrated with various $\Omega_{t}$. Then, with this transfer function the scattered optical power $P_{s}$ at $\Omega_{s}$ in the experiment can be deduced from the measured microwave power $P_{\mathrm{SA}}(\omega_{s})$ at $\omega_{s}$ with the spectrum analyzer by using the inverse relation
\begin{equation}
P_{s}(\Omega_{s}) = T(\Omega_{s} \rightarrow \omega_{s})^{-1}P_{\mathrm{SA}}(\omega_{s}). \label{eq:relation}
\end{equation}

The typical power spectra of the beat signal are shown in Fig.~\ref{plot:PSD}.
The noise represented by the white background is considered to be generated by the slight angular deviation between the crystal axis and the direction of the static magnetic field and the imperfect extinction ratio of the polarizers. The each power of the signal (noise) $P_{\mathrm{SA}}(\omega_{s})$ is calculated by summing up the power spectral density over the yellow region shown in Fig.~\ref{plot:PSD} around the beat frequency $\omega_{s}$.
Note that the finite linewidth of the signal can be attributed to the laser phase noise and optical path length fluctuation in addition to the filter bandwidth of the spectrum analyzer. The scattered photon flux $F_{s}(\Omega_{s})$ is deduced from the measured $P_{\mathrm{SA}}(\omega_{s})$ by using Eqs.~(\ref{eq:relation1}) and (\ref{eq:relation}). To confirm the reproducibility, the measurement is repeated six times, and the average and standard deviation of the scattered photon flux $F_{s}(\Omega_{s})$ are evaluated. 


\subsection{Evaluation of $n$}
We can measure the reflection coefficient $S_{11}(\omega)$ of the microwave signal fed into the loop coil by the network analyzer shown in Figs.~2(a) and (b) of the main text. The microwave reflection spectrum can be analyzed by the input-output theory~\cite{Clerk2010} to deduce the number of magnons $n$ excited in the sphere. The reflection coefficient $S_{11}(\omega)$ is given by
\begin{equation}
S_{11}(\omega)= \frac{i(\omega-\omega_{K})-\frac{1}{2}(\kappa_{e}-\gamma_{K})}{i(\omega-\omega_{K})-\frac{1}{2}(\kappa_{e}+\gamma_{K})},
\end{equation}
where $\omega_{K}$ is the resonance angular frequency of the Kittel mode, $\gamma_{K}$ is the intrinsic energy dissipation rate, and $\kappa_\mathrm{ext}$ is the coupling rate between the microwave field out of (into) a one-dimensional transmission line and the Kittel mode. Note that the total linewidth of the Kittel mode is then given by $\gamma_{t}=\gamma_{K}+\kappa_{e}$. From the fitting shown as the red lines in Figs.~2(a) and (b), we obtain $\omega_{K}/2\pi=5.07\,\mathrm{GHz}$, $\kappa_{e}/2\pi=6.3\,\mathrm{MHz}$, and $\gamma_{K}/2\pi=2.3\,\mathrm{MHz}$ for $\bm{H}_\mathrm{ext}\parallel \langle100\rangle$ and $\omega_{K}/2\pi=5.21\,\mathrm{GHz}$, $\kappa_{e}/2\pi=6.3\,\mathrm{MHz}$, and $\gamma_{K}/2\pi=2.0\,\mathrm{MHz}$ for $\bm{H}_\mathrm{ext}\parallel \langle111\rangle$.
In the case of the resonant excitation ($\omega=\omega_{K}$) the number of magnons $n$ reads 
\begin{equation}
n = \frac{4\kappa_{e}}{\gamma_{t}^2}\frac{P_\mathrm{mw}}{\hbar\omega},
\end{equation} 
where $P_\mathrm{mw}$ is the microwave power used to excite the magnons.
Thus the relation between the known microwave power $P_\mathrm{mw}$ and the number of magnon in the Kittel mode, $n$, is established.

\begin{figure}[b]
\includegraphics[width=8.6cm,angle=0]{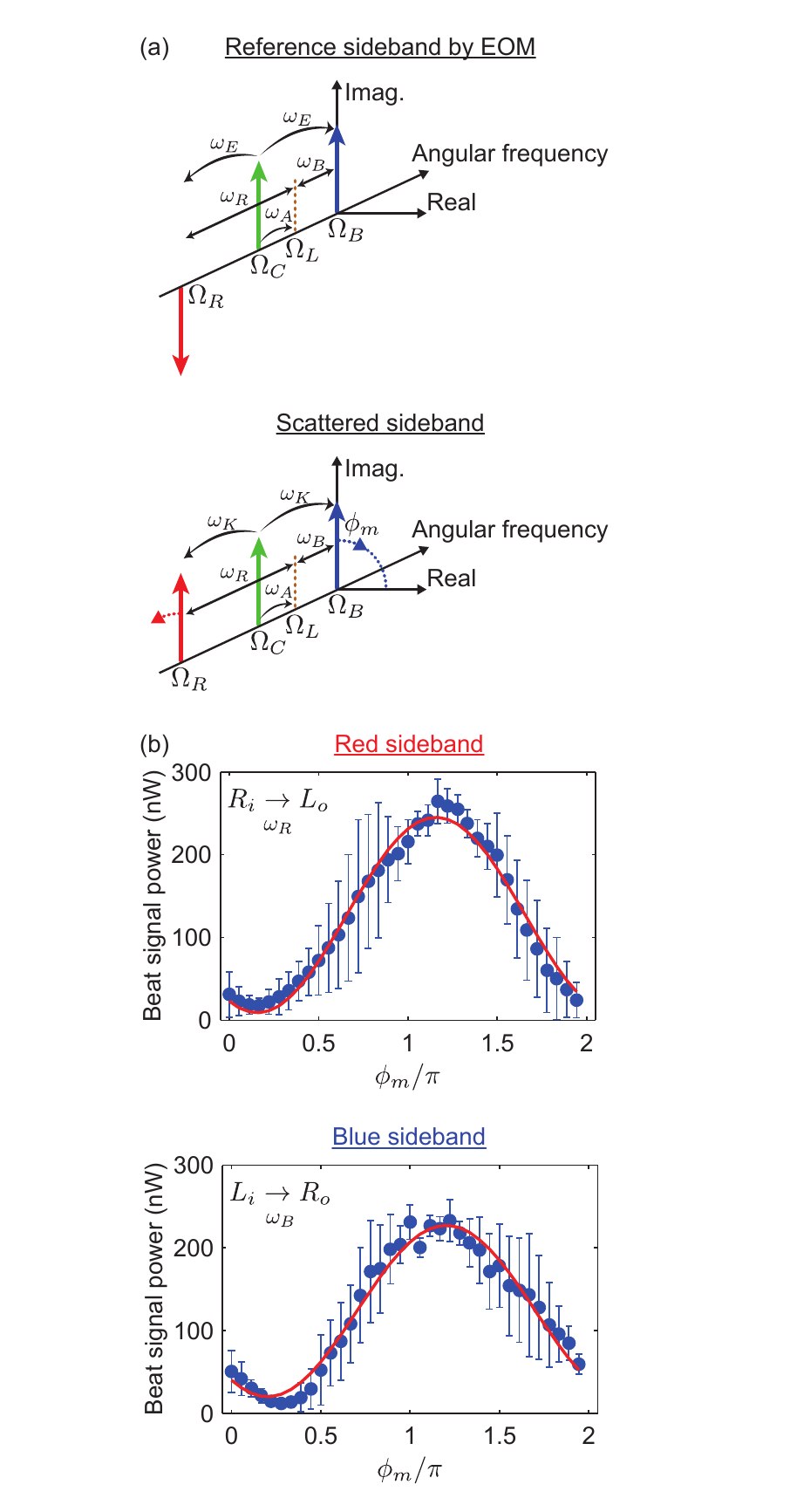}
\caption{(a) Schematics of the relevant frequencies in the experiment. Upper panel: The reference sideband at $\Omega_{R}$ and $\Omega_{B}$ are generated by the EOM driven at $\omega_{E}$ $(=\omega_{K})$. Here, the reference sideband has a fixed phase with respect to the carrier. Lower panel: The carrier light at $\Omega_{C}$ is scattered into the sidebands at $\Omega_{R}$ and $\Omega_{B}$ due to helicity-changing one-magnon scattering. Here, the phase of the scattered sideband with respect to the carrier varies as the phase $\phi_{m}$ of the magnon drive at $\omega_{K}$ varies. 
 (b) Observed interference fringes. Upper panel: The result of the beat signals (${R}_{i}\rightarrow{L}_{o}$ configuration) at $\omega_{R}$ originated from red sideband at $\Omega_{R}$ with the LO at $\Omega_{L}$.  
Lower panel: The result of the beat signals (${L}_{i}\rightarrow{R}_{o}$ configuration) at $\omega_{B}$ originated from blue sideband at $\Omega_{B}$ with the LO at $\Omega_{L}$.
}
\label{plot:coherence}
\end{figure}

\section{Experiments on coherence}
In this section, we experimentally check whether the helicity-changing one-magnon scattering assisted by crystal angular momentum for the case with $\bm{H}_{\mathrm{ext}}\parallel\langle111\rangle$ [Fig.~2(f)] preserves the phase coherence between the carrier light at $\Omega_{C}$ and the scattered sideband at $\Omega_{R}$ (or $\Omega_{B}$).
The basic idea to check this is to use another sideband at $\Omega_{R}$ (or $\Omega_{B}$) as a \textit{reference sideband} which is generated by the EOM shown in Fig.~1(b). The reference sideband has a fixed phase with respect to the carrier as shown in the upper panel in Fig.~\ref{plot:coherence}(a). The phase of the sideband generated by the helicity-changing one-magnon-scattering can be dictated by the phase of the microwave used for driving Kittel mode at $\omega_{K}$. Note that this is only true if the intervention of the crystal angular momentum in the one-magnon-scattering do not spoil the phase relationship between the carrier and the sideband. The phase of the scattered sideband with respect to the carrier would thus varies as the phase $\phi_m$ of the magnon drive at $\omega_{K}$ varies as shown in the lower panel in Fig.~\ref{plot:coherence}(a). By letting the reference sideband and the scattered sideband interfere the degree of coherence between the carrier and the scattered sideband can be revealed as the interference fringes of the power as a function of $\phi_m$.

The upper (lower) panel in Fig.~\ref{plot:coherence}(b) shows the observed interference fringes of the power for the red (blue) sideband at $\Omega_{R}$ ($\Omega_{B}$) [which appears at the beat angular frequency at $\omega_{R}$ ($\omega_{B}$) with respect to the LO light at $\Omega_{L}$]. To obtain the maximum interference fringes, (i) the polarization is set to be linearly polarized since the polarization of the reference sideband (same as that of the carrier) is opposite to that of the scattered sideband, and (ii) the scattered sideband power and reference sideband power are made equal at the HPD. The beat signal power is calculated by summing up the power spectral density for $\pm4$ Hz around the beat frequency. In order to check the reproducibility the measurement is repeated three times, and the average of the obtained signal power is shown by blue points, and the standard deviation is shown by blue bars. The red lines show the fitting results with a sinusoidal of period $2\pi$. The results thus prove that the intervention of the crystal angular momentum in the one-magnon Brillouin light scattering does not affect the coherence between the carrier and the scattered sidebands.


\begin{thebibliography}{99}

\bibitem{JS1987}
M.~Johnson and R.~H.~Silsbee, Phys.~Rev.~B~\textbf{35},~4959 (1987).

\bibitem{RS2008}
D.~C.~Ralph and M.~D.~Stiles, J.~Magn.~Magn.~Mater.~\textbf{320},~1190 (2008).

\bibitem{SVWBJ2015}
J.~Sinova, S.~O.~Valenzuela, J.~Wunderlich, C.~H.~Back, and T.~Jungwirth, Rev.~Mod.~Phys.~\textbf{87},~1213 (2015).


\bibitem{LL8}
L.~D.~Landau, E.~M.~Lifshitz, and L. P. Pitaevskii, \textit{Electrodynamics of Continuous Media}, 2nd ed. (Butterworth-Heinenann, Oxford, England, 1984).

\bibitem{Pershan}
P.~S.~Pershan, J.~Appl.~Phys.~\textbf{38}, 1482 (1967).

\bibitem{KKR2010}
A.~Kirilyuk, A.~V.~Kimel, and Th.~Rasing, Rev.~Mod.~Phys. \textbf{82,} 2731 (2010).

\bibitem{Osada2016}
A.~Osada, R.~Hisatomi, A.~Noguchi, Y.~Tabuchi, R.~Yamazaki, K.~Usami, M.~Sadgrove, R.~Yalla, M.~Nomura, and Y.~Nakamura, Phys.~Rev.~Lett.~{\bf 116}, 223601 (2016).

\bibitem{Zhang2016}
X.~Zhang, N.~Zhu, C.~-L.~Zou, and H.~X.~Tang, Phys.~Rev.~Lett.~{\bf~117}, 123605 (2016).

\bibitem{Haigh2016}
J. ~A.~Haigh, A.~Nunnenkamp, A. ~J.~Ramsay, and A.~ J.~Ferguson, Phys.~Rev.~Lett.~{\bf 117}, 133602 (2016).

\bibitem{Kusminskiy2016}
S.~ViolaKusminskiy, H.~X.~Tang, and F.~Marquardt, Phys.~Rev.~A~{\bf~94}, 033821 (2016).

\bibitem{Sharma2017}
S.~Sharma, Y.~M.~Blanter, and G.~E.~W.~Bauer, Phys.~Rev.~B~{\bf 96}, 094412 (2017).

\bibitem{Osada2018}
A.~Osada, A.~Gloppe, R.~Hisatomi, A.~Noguchi, R.~Yamazaki, M.~Nomura, Y.~Nakamura, and K.~Usami, Phys.~Rev.~Lett.~{\bf~120}, 133602 (2018).

\bibitem{Haigh2018}
J.~A.~Haigh, N.~J.~Lambert, S.~Sharma, Y.~M.~Blanter, G.~E.~W.~Bauer, and A.~J.~Ramsay, Phys.~Rev.~B~{\bf 97} 214423 (2018).

\bibitem{Osada2018_2} A.~Osada, A.~Gloppe, Y.~Nakamura, and K.~Usami, New~J.~Phys.~\textbf{20}, 103018 (2018).


\bibitem{FPCG1966}
P.~A.~Fleury, S.~P.~S.~Porto, L.~E.~Cheesman, and H.~J.~Guggenheim, Phys.~Rev.~Lett.~\textbf{17}, 84 (1966).

\bibitem{FPL1967}
P.~A.~Fleury, S.~P.~S.~Porto, and R.~Loudon, Phys.~Rev.~Lett.~\textbf{18}, 658 (1967).

\bibitem{Moriya1967}
T.~Moriya, J.~Phys.~Soc.~Jpn.~\textbf{23}, 490 (1967).

\bibitem{Moriya1968}
T.~Moriya, J.~Appl.~Phys.~\textbf{39}, 1042 (1968).

\bibitem{FL1968}
P.~A.~Fleury and R.~Loudon, Phys.~Rev.~\textbf{166}, 514 (1968).

\bibitem{Polzik1999}
J.~Hald, J.~L.~S{\o}rensen, C.~Schori, and E.~S.~Polzik, Phys.~Rev.~Lett.~\textbf{83}, 1319 (1999).

\bibitem{Oberthaler2011}
C.~Gross, H.~Strobel, E.~Nicklas, T.~Zibold, N.~Bar-Gill, G.~Kurizki, and M.~K.~Oberthaler, Nature~(London) \textbf{480}, 219 (2011).

\bibitem{Klempt2011}
B.~L{\"u}cke, M.~Scherer, J.~Kruse, L.~Pezz{\'e}, F.~Deuretzbacher, P.~Hyllus, O.~Topic, J.~Peise, W.~Ertmer, J.~Arlt, L.~Santos, A.~Smerzi, and C.~Klempt, Science~\textbf{334}, 773 (2011).

\bibitem{Chapman2012}
C.~D.~Hamley, C.~S.~Gerving, T.~M.~Hoang, E.~M.~Bookjans, and M.~S.~Chapman, Nat.~Phys.~\textbf{8}, 305 (2012).

\bibitem{KU1993}
M.~Kitagawa and M.~Ueda, Phys.~Rev.~A~\textbf{47}, 5138 (1993).



\bibitem{SB1968}
H.~J.~Simon and N.~Bloembergen, Phys.~Rev.~\textbf{171}, 1104 (1968).

\bibitem{Bloembergen1980}
N.~Bloembergen, J.~Opt.~Soc.~Am.~\textbf{70}, 1429 (1980).

\bibitem{Nienhuis2002}
J.~Visser, E.~R.~Eliel, and G.~Nienhuis, Phys.~Rev.~A~\textbf{66}, 033814 (2002).

\bibitem{Higuchi2011}
T.~Higuchi, N.~Kanda, H.~Tamaru, and M.~Kuwata-Gonokami, Phys.~Rev.~Lett.~\textbf{106}, 047401 (2011).

\bibitem{Konishi2014}
K.~Konishi, T.~Higuchi, J.~Li, J.~Larsson, S.~Ishii, and M.~Kuwata-Gonokami, Phys.~Rev.~Lett.~\textbf{112}, 135502 (2014).

\bibitem{GM}
A.~G.~Gurevich and G.~A.~Melkov, \textit{Magnetization Oscillations and Waves} (CRC Press, Boca Raton, 1996).

\bibitem{Stancil}
D.~D.~Stancil and A.~Prabhakar, \textit{Spin Waves: Theory and Applications} (Springer, New York, 2009).

\bibitem{SM}
Appendix, which includes Refs.~\cite{BKD,LL7,Nye,HM1967,Mabuchi2006,Polzik2010,Kittel,CDG,Healy,Clerk2010} and details on the theory, the determination of the crystal axis, the calibration of scattering efficiencies, and experiments on coherence.

\bibitem{BKD}
D.~Budker, D.~F.~J.~Kimball, and D.~DeMille, \textit{Atomic Physics}, 2nd ed. (Oxford University Press, Oxford, England, 2008).

\bibitem{LL7}
L.~D.~Landau and E.~M.~Lifshitz, \textit{Theory of Elasticity}, 3rd ed. (Butterworth-Heinenann, Oxford, England, 1986).

\bibitem{Nye}
J.~F.~Nye, \textit{Physical Properties of Crystals} (Oxford University Press, Oxford, 1985).

\bibitem{HM1967}
W.~Happer and B.~S.~Mathur, Phys.~Rev.~\textbf{163}, 12 (1967).

\bibitem{Mabuchi2006}
J.~M.~Geremia, J.~K.~Stockton, and H.~Mabuchi, Phys.~Rev.~A~\textbf{73}, 042112 (2006).

\bibitem{Polzik2010}
K.~Hammerer, A.~S.~S{\o}rensen, and E.~S.~Polzik, Rev.~Mod.~Phys.~\textbf{82}, 1041 (2010).

\bibitem{Kittel}
C.~Kittel, \textit{The Quantum Theory of Solids} (Wiley, New York, 1963).

\bibitem{CDG}
C.~Cohen-Tannoudji, J.~Dupont-Roc, and G.~Grynberg, \textit{Atom-Photon Interactions} (John Wiley and Sons, New York, 1992).

\bibitem{Healy}
D.~Healy Jr., Phys.~Rev.~\textbf{86}, 1009 (1952).

\bibitem{Clerk2010}
A.~A.~Clerk, M.~H.~Devoret, S.~M.~Girvin, F.~Marquardt, and R.~J.~Schoelkopf,  Rev.~Mod.~Phys. \textbf{82,} 1155 (2010).

\bibitem{Tinkham}
M.~Tinkham, \textit{Group Theory and Quantum Mechanics} (McGraw-Hill, New York, 1964).

\bibitem{HL}
W.~Hayes and R.~Loudon, \textit{Scattering of Light by Crystals} (Wiley, New York, 1978).

\bibitem{Budker2019}
T.~Wang, S.~Lourette, S.~R.~OKelley, M.~Kayci, Y.~B.~Band, Derek~F.~Jackson~Kimball, A.~O.~Sushkov, and D.~Budker,  	Phys.~Rev.~Applied~\textbf{11}, 044041 (2019).

\bibitem{Hetet2019}
P. Huillery, T. Delord, L. Nicolas, M. Van Den Bossche, M. Perdriat, and G. H\'etet, arXiv:1903.09699


\end{thebibliography}
\end{document}